\documentstyle[erice,12pt,epsfig]{article}

\newcommand{\be}[1]{\begin{equation} \label{(#1)}}
\newcommand{\ee}{\end{equation}}
\newcommand{\ba}[1]{\begin{eqnarray} \label{(#1)}}
\newcommand{\ea}{\end{eqnarray}}

\begin{document}  
\title{Dileptons, Charm and Bottom in Relativistic Heavy-Ion Collisions}
\author{B. K\"ampfer$^a$, O.P. Pavlenko$^{a,b}$, K. Gallmeister$^a$}
\address{$^a$Forschungszentrum Rossendorf, PF 510119, 01314 Dresden, Germany\\
$^b$Institute for Theoretical Physics, 252143 Kiev - 143, Ukraine}
\maketitle
\abstracts{
The relation of the thermal dilepton signal from deconfined matter,
resulting in central ultra-relativistic heavy-ion collisions at RHIC and
LHC energies, to the background yields from the Drell-Yan process 
and correlated
semileptonic decays of open charm and bottom mesons is analyzed.
We demonstrate that very stringent kinematical cuts offer a chance to identify
the thermal signal in the continuum region.}

\section{Introduction}

Since a long time the penetrating probes, such as dileptons and
photons, are considered as nearly undisturbed messengers from hot and dense,
strongly interacting matter produced in central ultra-relativistic
heavy-ion collisions (cf. \cite{Ruusk}).
Indeed, in present SPS experiments
($\sqrt{s} = 15 \cdots 20$ GeV) the CERES collaboration reports a dilepton
excess over the known hadronic cocktail in S + Au and Pb + Au collisions
at invariant masses $M < 1$ GeV \cite{CERES-QM}.
The order of magnitude of this excess
can be attributed to a thermal source stemming mainly from pion annihilation,
while the detailed shape of the spectrum is still matter of debate,
e.g. it might reflect an in-medium changed $\rho$ spectral function.
Similarly, in the so-called intermediate mass region 
$M = M_\phi \cdots M_{J/\psi}$, as accessible in the acceptances of the
HELIOS-3 and NA38/50 experimental set-ups, the conventional sources
Drell-Yan (DY) and open charm decays seem also not to account for the
observed data, i.e. there is also an excess \cite{excess}. 
It is just the intermediate mass region where all model calculations 
(cf. \cite{Ruusk, PRC95}) predict
best chances to see a thermal signal from deconfined matter. 

For an illustration of that fact we display in Fig.~1 the time evolution
of the thermal dilepton yield from purely deconfined matter for
various invariant masses covering the intermediate mass region.
Indeed,
as seen in Fig.~1, the major amount (90\%) of harder dileptons with
$M = 4$ GeV stem from early ($\tau = 0.2 - 2$ fm/c)
and hot ($T > 2 T_c$) matter stages, while the more soft dileptons
with $M = 1$ GeV are breeded out mainly at temperatures around
confinement temperature $T_c$ and are strongly contaminated by dileptons
from the hadron stage.

Analyzes of hadron particle ratios \cite{ratios} 
and transverse momentum spectra \cite{BK} in present CERN-SPS experiments
deliver temperatures of $T \sim {\cal O}(m_\pi)$. While this is a first
step to pin down the production of a hot system in heavy-ion collisions,
one is interested in signals which can identify much higher temperatures
pointing to the necessary achievement of conditions for deconfinement,
i.e. $T \gg {\cal O}(m_\pi)$. The measured hadron rapidity densities
at SPS can be translated into maximum initial temperatures
$T_i \sim {\cal O}(200)$ MeV when using Bjorken's estimate with
a initial time $\tau_i = 1$ fm/c.

At higher beam energies one expects also much higher maximum temperatures
achievable. On the same time, when going to higher center-of-mass
energies $\sqrt{s}$, also the background from DY and the hadronic decay
cocktail increases. In particular, correlated semileptonic decays of open charm
and bottom constitute a severe background.
In what follows we consider the beam energy dependence of the dilepton
yield in the intermediate mass region
(Section 2). It turns out that the mentioned background
processes are much stronger than the thermal signal. We therefore study then
in Section 3 various kinematical cuts, which suppress the background, and
analyze the possibility to identify the thermal signal from deconfined
matter. We focus here on this signal, since the dilepton radiation
from hadron matter or a possible mixed phase will again deliver only a hint
to relatively cold matter with $T \sim {\cal O}(m_\pi)$.
Our conclusions can be found in Section 4.
\begin{figure}[t]
\begin{minipage}[t]{7.5cm}
\centering
~\\[-.3cm]
\psfig{file=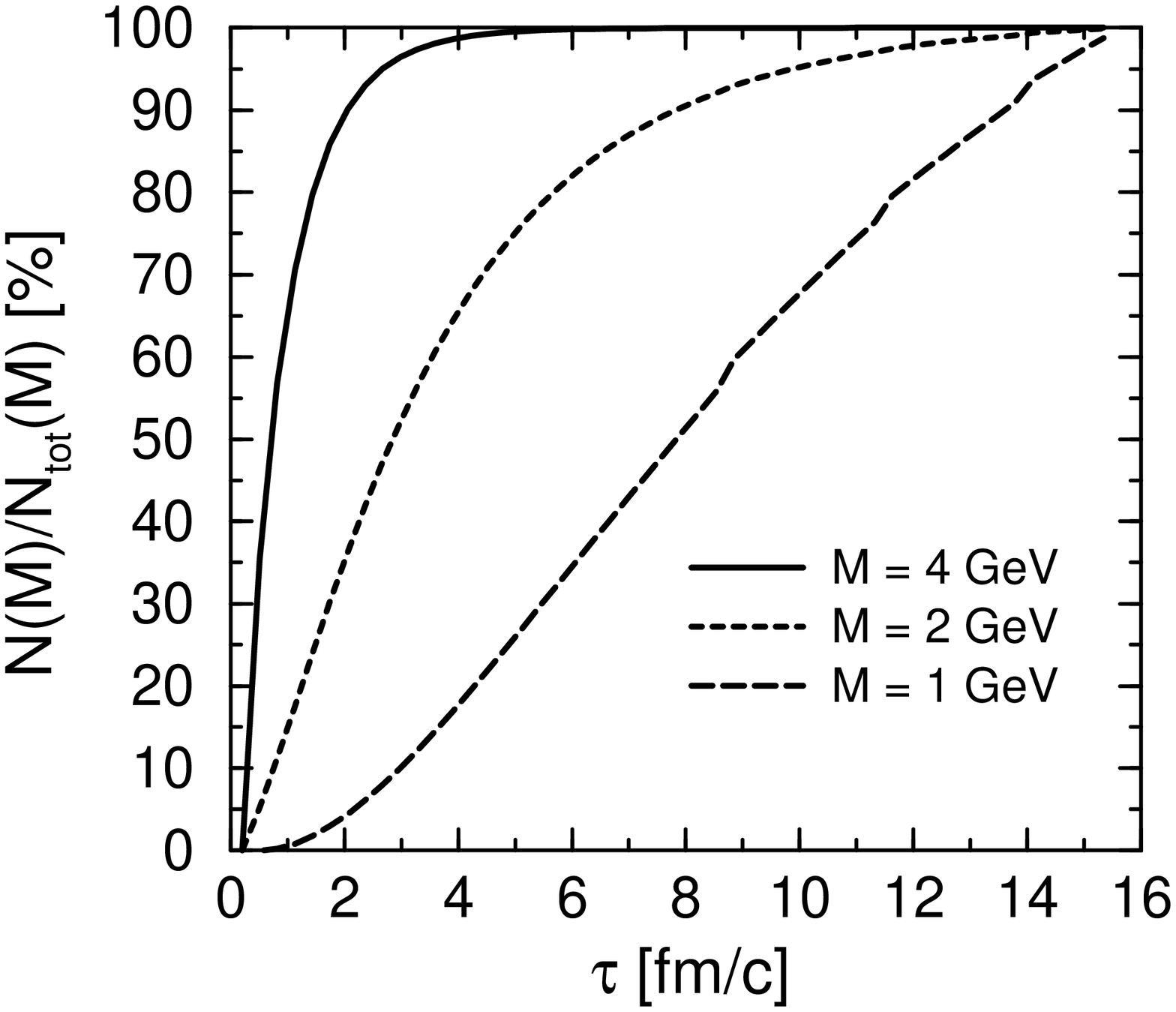,width=6.5cm,angle=-90}
~\\[.01cm]
\caption{
The time dependence of the normalized dilepton yield from purely
deconfined matter
at midrapidity for
various invariant masses and for LHC initial conditions.}
\label{fig.1}
\end{minipage}
\hspace*{1cm}
\begin{minipage}[t]{7.5cm}
\centering
~\\[-.1cm]
\psfig{file=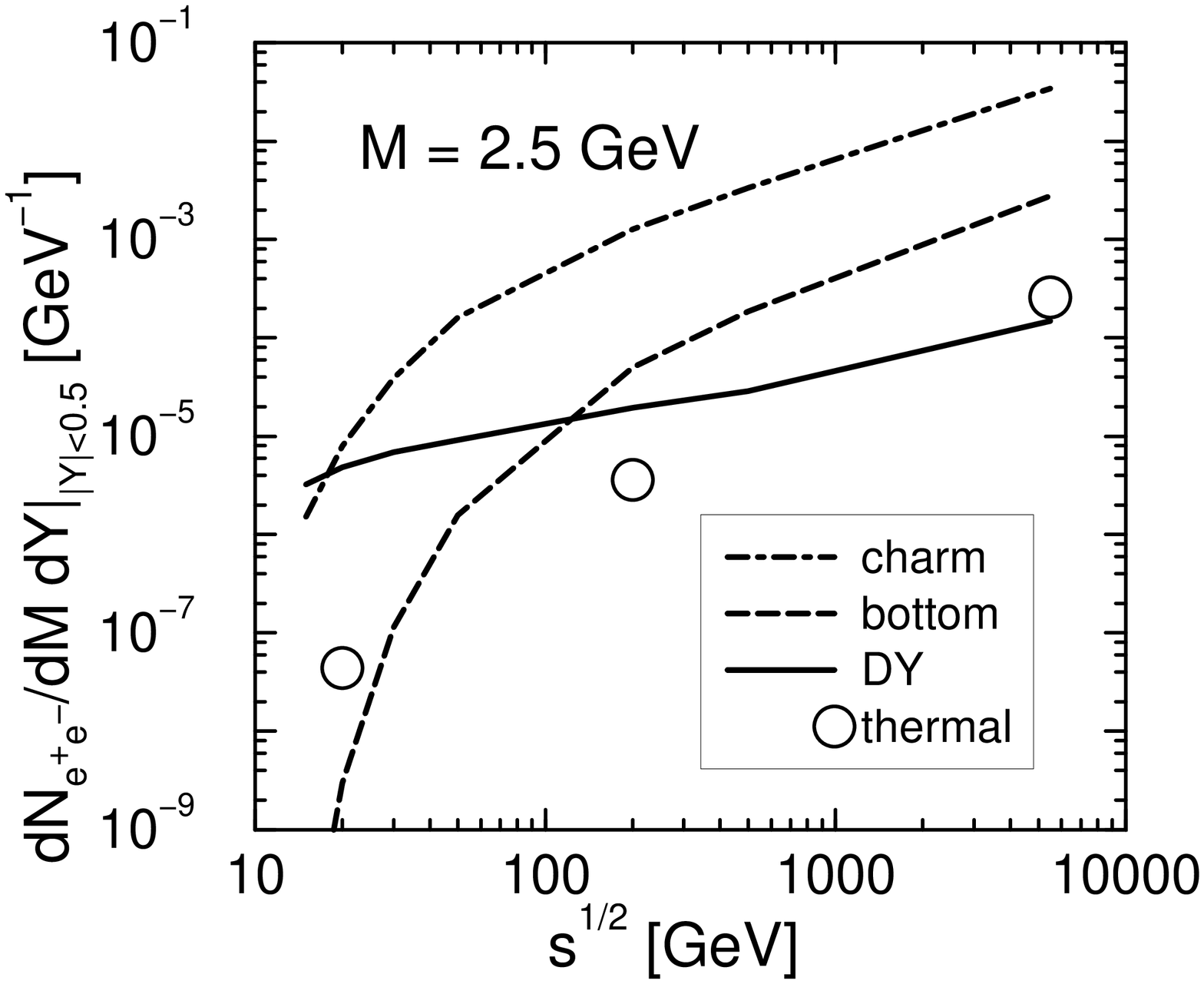,width=6.3cm,angle=-90}
~\\[.01cm]
\caption{The dependence of dileptons from the lowest-order
processes (DY and correlated semileptonic decays of open charm or
bottom mesons)
and the thermal source on $s^{1/2}$.}
\label{fig.2}
\end{minipage}
\end{figure}

\section{Beam energy dependence}

The necessary formulae for calculating the mentioned dilepton yields
within a lowest-order approach are accumulated in \cite{PRC98,FZR-235}.

The initial conditions of the thermal era are estimated within a mini-jet
picture. It allows to calculate the energy distribution and number
distribution of mini-jets. Transport models and event generators point to
the fact that this mini-jet plasma is thermalized at a very early time
instant, $\tau_i \sim 0.2$ fm/c. The resulting values of initial temperatures
$T_i$, gluon fugacities $\lambda_i^g$ and resulting hadron rapidity densities
$dN_\pi/dy$ are listed in Tab.~1 \cite{PLB97}.
The corresponding quark fugacities are
usually $\lambda_i^q = \frac 15 \lambda_i^g$.\\[6mm] 
\begin{minipage}[t]{7cm}
\begin{center}
\begin{tabular}{|l||c|c|}
\hline
 & RHIC & LHC   \protect \\ 
\hline \hline
$T_i$ [GeV]  & 0.544 & 1.038 \protect\\
\hline
$\lambda_i^g$& 0.41  &0.25    \protect\\
\hline
$dN_\pi/dy$&1080&5204  \protect\\
\hline
\end{tabular}
\end{center}
\end{minipage}
\hspace*{1cm}
\begin{minipage}[t]{7cm}

\vspace*{-1.5cm}

Table 1: Estimates of initial values achievable at RHIC and LHC
within the mini-jet picture. Nuclear shadowing effects are included.
\end{minipage}

\vspace*{8mm}

With these propositions one gets
the beam energy dependence of dileptons with $M \sim 2.5$ GeV at midrapidity
as displayed in Fig.~2.
The thermal source includes only purely
deconfined matter.
Note that at SPS energies the hadron and a possible mixed phase
(both ones not included here) represent also strong thermal sources.
One observes that the ratio of DY dileptons to correlated open charm (bottom)
decay dileptons drops down with increasing
$\sqrt{s}$. Only in the region $\sqrt{s} <$ 20 (120) GeV the
Drell-Yan yield dominates above the charm (bottom) decay contribution.
Note that the thermal yield estimates depend sensitively on the assumed
initial parameters.
In agreement with recent findings the thermal dilepton
signal is up to two orders of magnitude below the correlated open charm
decay dileptons in a wide range of invariant masses at LHC,
RHIC \cite{Vogt2} and SPS energies.
Therefore, there is nowhere a preferred energy region;
at high beam energies, however, one expects clearer deconfinement effects
due to temperatures far above the confinement temperature.

The results displayed in Fig.~2 cover
the full phase space. Any detector acceptance will suppress the
various sources differently \cite{PRC98}. Energy loss of heavy quarks
propagating through deconfined matter reduces also the open charm and bottom
decay yields \cite{Shur1_Lin2,PLB98}.
Notice that with increasing invariant mass the thermal yield drops
exponentially while, for instance, the Drell-Yan yields drops less, i.e.
like a power law.
Therefore, the results displayed in Fig.~2 can not be extrapolated in
a simple manner to higher invariant masses.

\section{Kinematical cuts}

From all of these considerations the problem arises whether one
can find such kinematical gates which enable one to discriminate the thermal
signal against the large decay background. Since the kinematics of
heavy meson production and decay differs from that of thermal dileptons,
one can expect that special kinematical restrictions superimposed on
the detector acceptance will be useful for finding a window for
observing thermal dileptons in the intermediate mass continuum region.
In particular, experimental cuts on the rapidity gap between the leptons
can reduce considerably the charm decay background \cite{Vogt2}.
As demonstrated recently \cite{PRC98}, the measurement of the double
differential dilepton spectra as a function of the transverse pair
momentum $Q_\perp$ and transverse mass $M_\perp = \sqrt{M^2 + Q_\perp^2}$
within a narrow interval of $M_\perp$ also offers the chance to observe
thermal dileptons at LHC.
In the present note we show that a large enough low-$p_\perp$ cut on
single electrons suppresses the mentioned background processes and
opens a window for the thermal signal
in the invariant mass distribution. We take into account
the acceptance of the ALICE detector at LHC: the single electron
pseudo-rapidity is limited to $\vert \eta_e \vert \le 0.9$ and an
overall low-$p_\perp$ cut of 1 GeV is applied.
We are going to study systematically the effect of enlarging the
low-$p_\perp$ cut on single electrons in the invariant-mass
dilepton spectra at $\sqrt{s} =$ 5500 GeV.

Since the energy of individual decay electrons or positrons has a
maximum of about 0.88 (2.2) GeV in the rest frame of the decaying
$D$ ($B$) meson, one can expect to get a strong suppression of correlated
decay lepton pairs by choosing a high enough low-momentum cut
$p_\perp^{\rm min}$ on the individual leptons in the mid-rapidity region.
For thermal leptons stemming from deconfined matter there is no such upper
energy limit and for high temperature the thermal yield will not suffer
such a drastically suppression by the $p_\perp^{\rm min}$ cut as the
decay background.
To quantify this effect we perform Monte Carlo calculations generating
dileptons from the sources mentioned above. We first employ the
transparent lowest-order calculations and then present 
more complete PYTHIA simulations. 

As heavy quark masses we take $m_c =$ 1.5 GeV and $m_b =$ 4.5 GeV.
We employ the HERA supported structure function set MRS D-' \cite{MRS}
from the PDFLIB at CERN.
Nuclear shadowing effects are not needed to be included for our present
order of magnitude estimates.

The overlap function for central collisions is
$T_{AA}(0) = A^2/(\pi R_A^2)$
with  $R_A = 1.2 A^{1/3}$ fm and $A = 200$.
From a comparison with results of Ref.~\cite{Vogt2}
we find the scale $\hat Q^2 = 4 m_Q^2$ and ${\cal K}_Q =$ 2 as most
appropriate.

We employ a delta function fragmentation scheme for the heavy quark
conversion into $D$ and $B$ mesons.
Dilepton spectra from correlated semileptonic decays,
i.e., $D (B) \bar D (\bar B) \to e^+ X e^- \bar X$,
are obtained from a Monte Carlo code which utilizes the inclusive
primary electron energy distribution as delivered by
the JETSET part of PYTHIA 6.1. The heavy mesons are randomly decayed in their rest system
and the resulting electrons then boosted appropriately.
The average branching ratio of $D (B) \to e^+ X$
is taken as 12 (10)\%. 
\begin{figure}[t]
\begin{minipage}[t]{7.5cm}
\centering
~\\[-.1cm]
\psfig{file=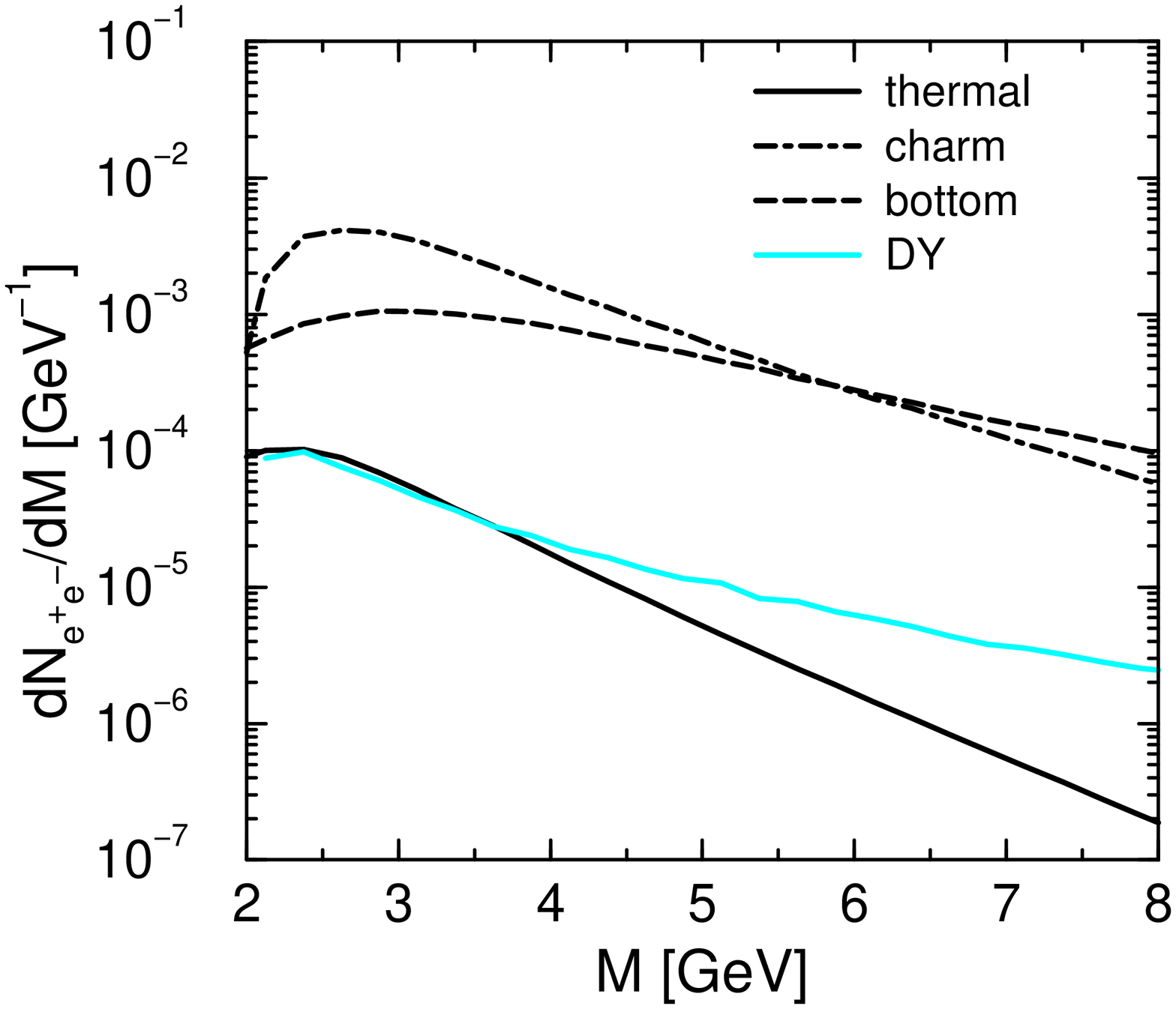,width=6.5cm,angle=-90}
~\\[.1cm]
\caption{The invariant mass spectra of dileptons from the
DY process, charm and bottom decays, and thermal emission.
The single-electron low transverse momentum cut is
$p_\perp^{\rm min} =$ 1 GeV.}
\label{fig.3}
\end{minipage}
\hspace*{1cm}
\begin{minipage}[t]{7.5cm}
\centering
~\\[-.1cm]
\psfig{file=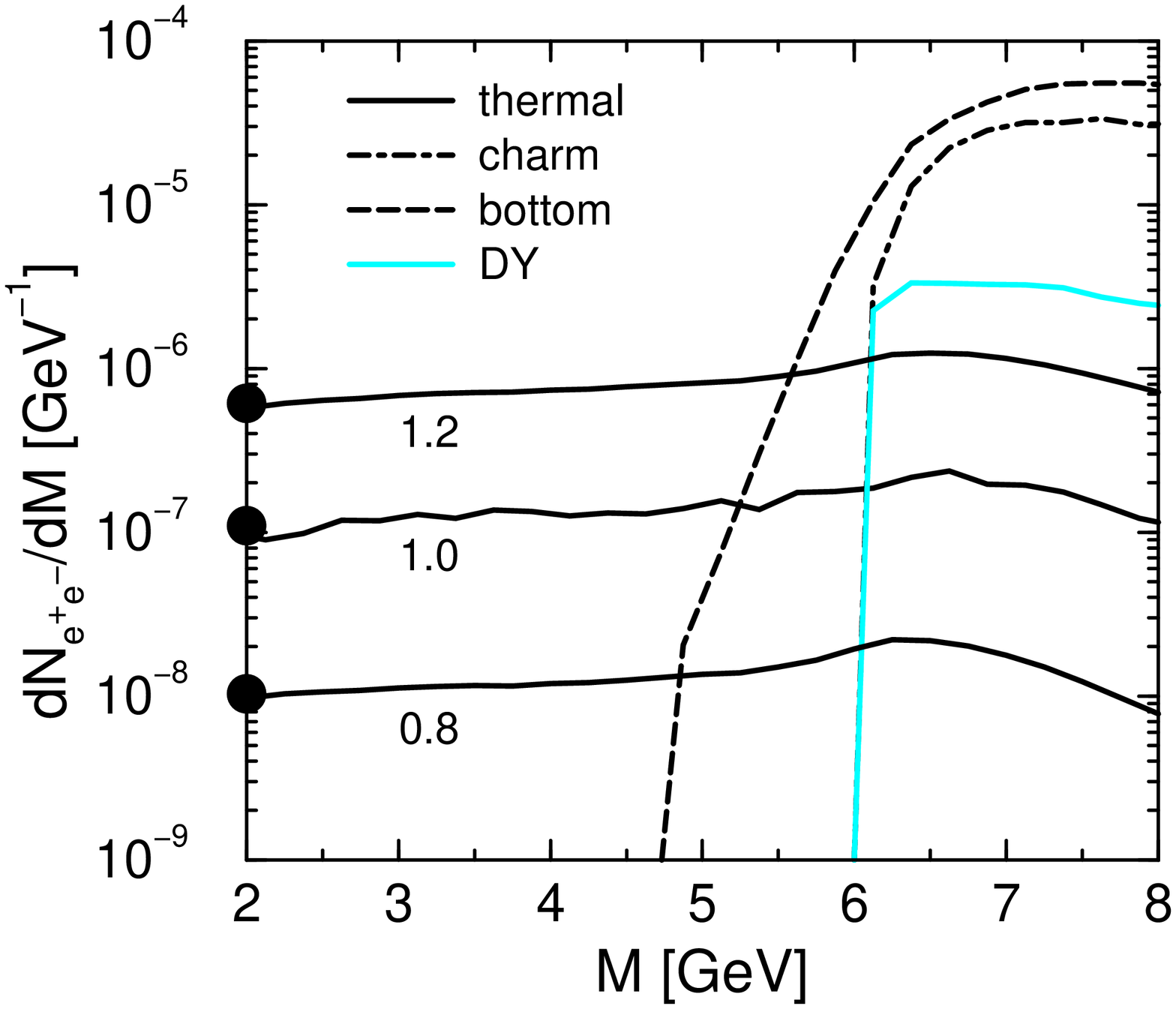,width=6.5cm,angle=-90}
~\\[.1cm]
\caption{The same as in Fig.~3, but
$p_\perp^{\rm min} =$ 3 GeV.
The fat dots indicate the estimates of the low-$M$ thermal 
plateau as described in \protect\cite{FZR-235}.}
\label{fig.4}
\end{minipage}
\end{figure}

The results of our lowest-order calculations of the invariant mass spectrum
for various values of $p_\perp^{\rm min}$ are displayed in Figs.~3 and
4 for LHC energies. One observes that 
the thermal dilepton signal with a single-electron low-momentum cut-off
$p_\perp^{\rm min} =$ 3 GeV exhibits an approximate plateau 
in the invariant mass region 2 GeV $\le M \le 2 p_\perp^{\rm min}$.
The physical information encoded in the continuum spectrum
is discussed in \cite{Gallmeister}.
The main point is that the cut
$p_\perp^{\rm min} =$ 3 GeV causes a strong suppression of the correlated
charm decay and DY background in the region $M \le 2 p_\perp^{\rm min}$.
The above value of the invariant mass threshold can be estimated by using
the relation
$M^2 =2 p_{\perp+} p_{\perp-} [\mbox{ch} (y_+ -y_-) - \cos(\phi_+ - \phi_-)]$,
where $\phi_{\pm}$ denote the azimuthal angles of the leptons
in the transverse plane.
In order to exceed the cut $p_\perp^{\rm min}$ most easily, the decay
leptons should go parallel to the parent heavy mesons, which in turn
are back-to-back (in the transverse plane)
in lowest order processes. As a consequence,
$\cos(\phi_1 - \phi_2) \approx -1$ and 
the minimum invariant mass becomes 
$M^{\rm min} \approx 2 p_\perp^{\rm min}$
for such decay pairs. For correlated bottom decay the electron energy
is larger in the meson rest system and both leptons can more easily overcome
the threshold $p_\perp^{\rm min}$ without such strong back-to-back correlation.
Selecting, however, electrons with $p_\perp > p_\perp^{\rm min} =$ 3 GeV one can
also get the corresponding threshold like behavior for the lepton pairs
from correlated bottom decay as seen in Fig.~4.
Therefore the thermal signal becomes clearly visible for such a value
of $p_\perp^{\rm min}$ due to the strong suppression of the considered
background channels.

The threshold behavior does not change if we include in our calculations
energy loss effects of heavy quarks in deconfined matter.
Such effects cause mainly a suppression of the
decay contributions (cf. \cite{Shur1_Lin2,PLB98}).
We also mention
that shadowing effects, not included in our Drell-Yan and heavy quark
production estimates, will diminish these yields somewhat. 

The lowest order DY yield has anyway $M^{\rm min} = 2 p_\perp^{\rm min}$.
From the above given relations for $M$ and $M_\perp$ one can derive the
inequality
$M \ge 2 p_\perp^{\rm min}
\sqrt{1 - \left(\frac{Q_\perp}{2 p_\perp^{\rm min}}\right)^2}$.
This relation tells us that in the region $M <$ 5 (4) GeV only pairs with
total transverse momentum $Q_\perp >$ 3.25 (4.5) GeV can contribute if
$p_\perp^{min} =$ 3 GeV.
Since the next-to-leading order DY distribution
$dN/dM^2 dQ_\perp^2 dY$ drops from $Q_\perp \approx$ 0 to 3 - 4 GeV
by nearly three orders of magnitude \cite{HardProbes}
one can estimate a small
higher order DY contribution in the small-$M$ region.
To quantify the smearing of the threshold effect by an intrinsic
$p_\perp$ distribution of initial partons we perform simulations with
the event generator
PYTHIA (version 6.104 \cite{PYTHIA}) with default switches.
Results are displayed in Fig.~5 and
show that the sharp threshold
effect from the above lowest-order Drell-Yan process is indeed
somewhat smeared out,
however, the small-$M$ region is still clean.
It turns out that the initial state radiation of partons
before suffering a hard collision is the main reason for rising the
pair $Q_\perp$ and for smearing out the sharp threshold effect, while
the intrinsic $p_\perp$ distribution of partons
causes minor effects.
\begin{figure}[t]
\begin{minipage}[t]{7.5cm}
\centering
~\\[-.1cm]
\psfig{file=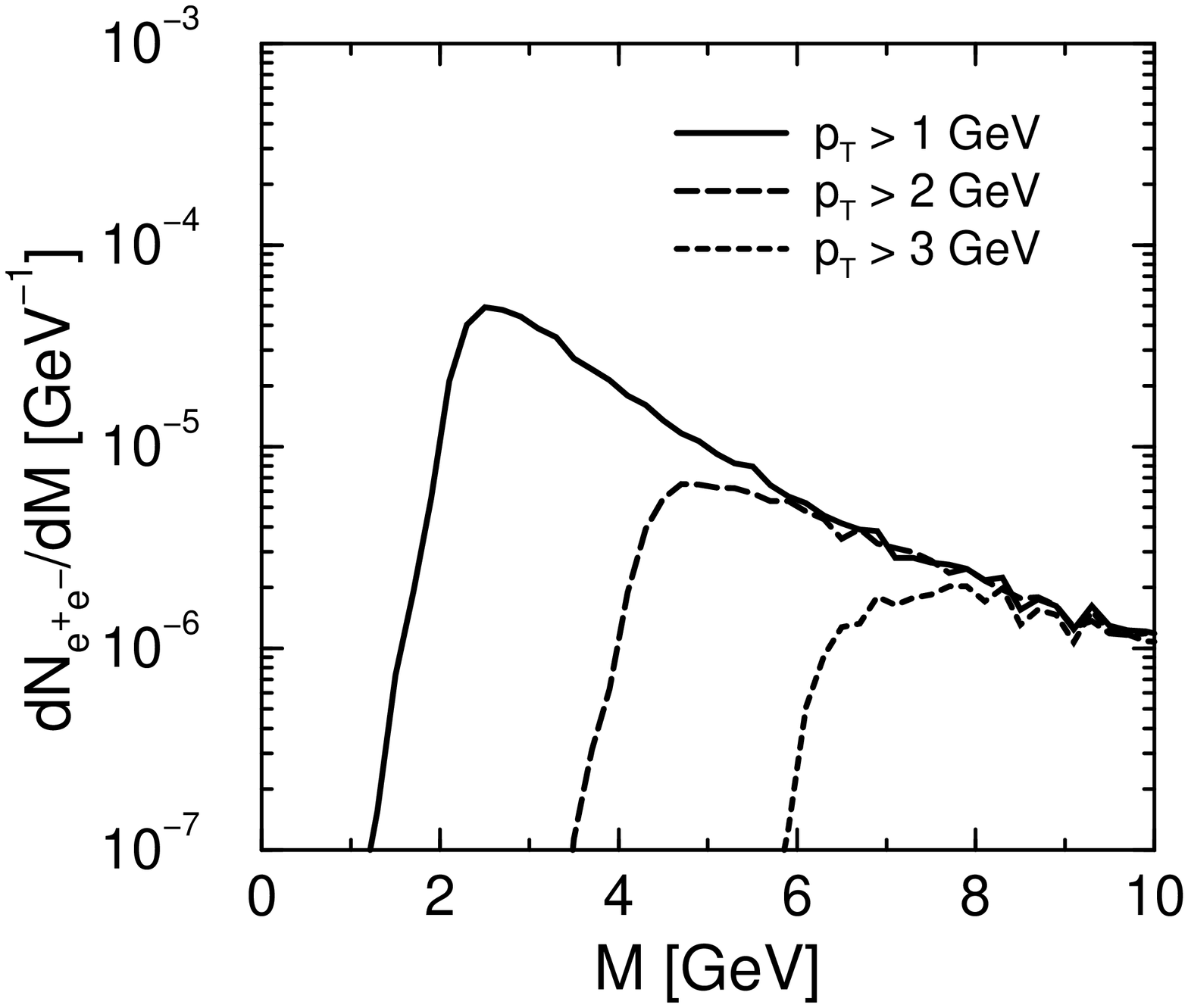,width=6.5cm,angle=-90}
~\\[.1cm]
\caption{The invariant mass spectra of dileptons from the
DY process for $p_\perp^{\rm min} =$
1, 2 and 3 GeV (from left to right). The curves depict results of 
PYTHIA with default switches.
}\label{fig.5}
\end{minipage}
\hspace*{1cm}
\begin{minipage}[t]{7.5cm}
\centering
~\\[-.1cm]
\psfig{file=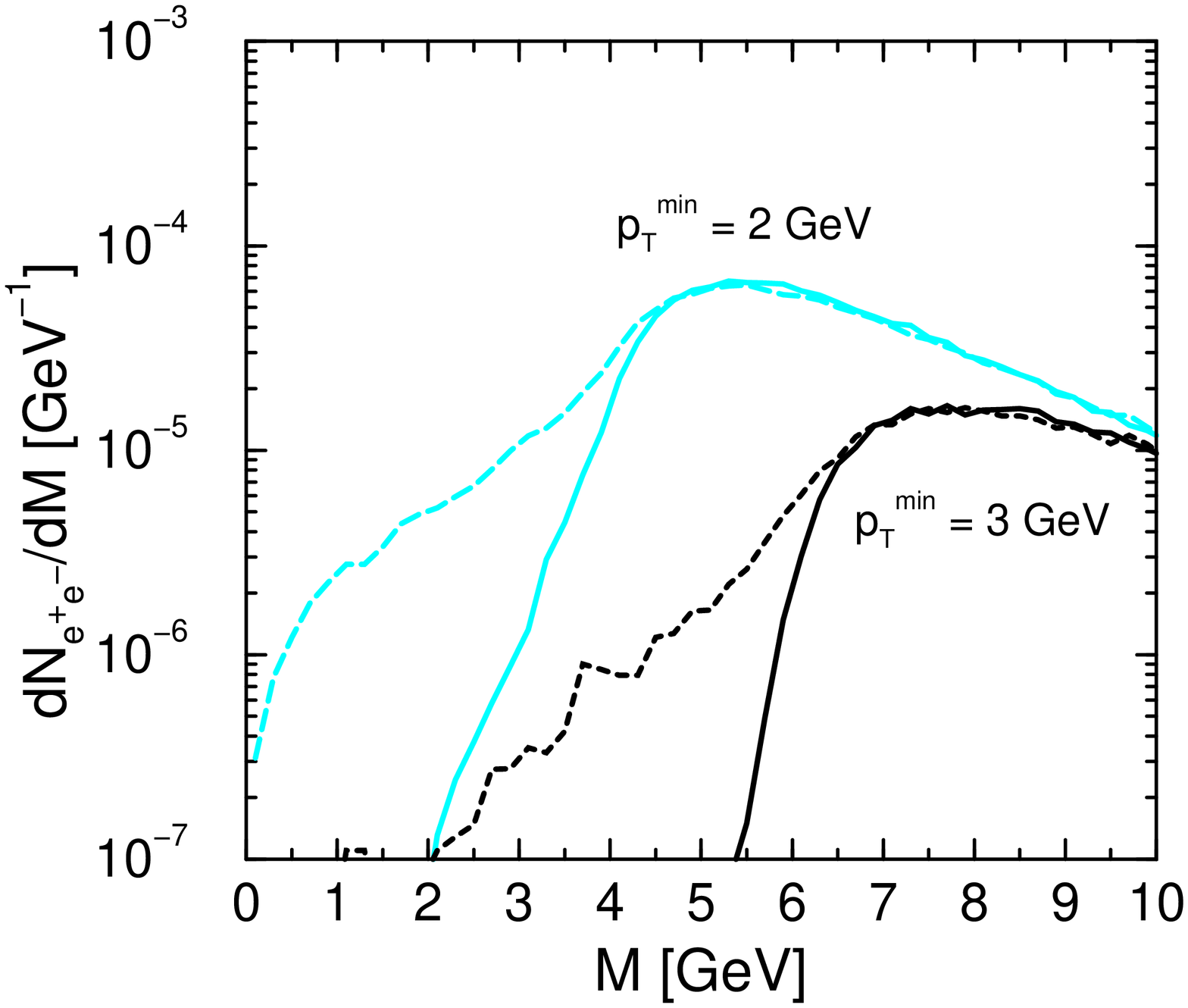,width=6.5cm,angle=-90}
~\\[.1cm]
\caption{The invariant mass spectra of dileptons from
correlated open bottom meson decays for $p_\perp^{\rm min} =$
2 and 3 GeV as delivered by PYTHIA 
(dashed curves: default switches, 
solid curves: without initial state radiation).
}
\label{fig.6}
\end{minipage}
\end{figure}

Let us now consider heavy quark pairs. Here,
the intrinsic $p_\perp$ distribution, and both initial and
final state radiations of the partons can cause a finite $Q_\perp$ and
thus destroy the strong back-to-back correlation,
i.e. $p_{\perp 1} \ne p_{\perp 2}$.
In Fig.~6 we show results of simulations with PYTHIA for the bottom
channel, where the resulting primary dileptons from all correlated open bottom
mesons are displayed. One observes that the initial state radiation
causes a pronounced smearing of the threshold effect discussed above.
Without the initial state radiation, as implemented in PYTHIA, the
threshold effect is recovered. The intrinsic $p_\perp$ distribution
and final state radiation are negligible. The conclusion of such
studies is that an enlarged value of $p_\perp^{\rm min}$ 
is necessary to keep
clean the low-$M$ region from open bottom decay products. For charm
the smearing effect due to initial state radiation in PYTHIA is
efficiently suppressed by the large enough low-$p_\perp$ cut of 
$p_\perp^{\rm min} =$ 3 GeV.
The different behavior of charm and bottom stem from the fact that the
bottom-$p_\perp$ distribution is much wider. As a consequence, the
bottom is evolved in the average to much larger values of 
the resolution scale $\hat Q^2$
thus experiencing stronger kicks by initial state radiation.
In agreement with our previous findings \cite{PLB98}, bottom therefore
causes the most severe background processes at LHC energies.

With PYTHIA also the dileptons from single decay chains of open bottom,
like $B^0 \to e^+ D^- (\to e^- \bar X) X$ or
$ B^0 \to e^+ D^*_{2010} (\to \bar D^{0,-} [\to e^- X'] \bar X) X$,
are accessible. These channels provides a contribution peaking at 1.5
GeV;
the cut $p_\perp^{\rm min} = 3$ GeV pushes all invariant masses below
3 GeV. Therefore, unless enlarging $p_\perp^{\rm min}$ considerably,
it will be difficult to suppress kinematically the background below the
thermal signal at invariant masses $M < 3$ GeV. Probably explicit
identification and subtraction of the bottom contribution is needed
in this region. 

Indeed, as recently found by the ALICE-GSI group \cite{PBM}, 
via exact tracking and vertex reconstruction one can suppress a substantial
part of the open charm and bottom decay electrons in the midrapidity
region. Therefore, the need of stringent cuts is relaxed somewhat and
realistic count rates are to be expected.

In this respect we would also like to point out that an explicit measurement
of the inclusive single-electron $p_\perp$-spectra 
from open charm and bottom decays contains valuable
information \cite{PRC98}. 
Namely the energy losses due to induced gluon radiation \cite{Baier}
of heavy quarks propagating through deconfined matter change the resulting
momentum distribution of the open charm and bottom mesons and, 
as a consequence,
the decay electrons exhibit a significantly changed $p_\perp$ spectrum
(for details consult \cite{PRC98}). Since such an effect does not appear in pp
collisions, the verification of a modified electron spectrum from identified
charm and bottom decays would offer a hint to the creation of deconfined
matter. The studies in \cite{PBM} demonstrate that tracking cuts offer
the chance to get a ''signal''-to-background ratio of 98\%, where
''signal'' means here the decay electrons from charm and bottom.
Therefore, such a measurement seems to be feasible with ALICE at LHC.
To illustrate the order of magnitude of the expected effect we show in
Fig.~7 the transverse momentum spectrum of decay electrons from open
charm and bottom mesons in a lowest-order calculation as described
above \cite{PRC98}.
One can fit the distribution by 
$dN_{e^-}/dp_\perp \propto \exp ( -p_\perp/T_e )$ in the interval
2.5 GeV $< p_\perp <$ 4.5 GeV and finds a change of the slope
parameter $T_e$ from 850 MeV (without energy loss) to
640 MeV (with energy loss).
\begin{figure}[t]
\begin{minipage}[t]{10cm}
\centering
~\\[-.1cm]
\psfig{file=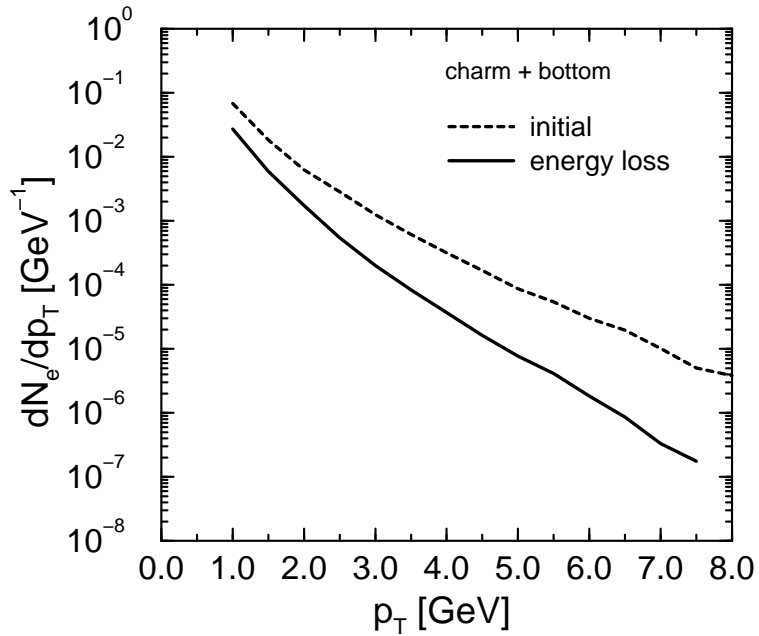,width=8.5cm,angle=-90}
~\\[.1cm]
\label{fig.7}
\end{minipage}
\hspace*{1cm}
\begin{minipage}[t]{5cm}
\centering
~\\[6mm]
\caption{The transverse momentum spectra of single electrons from $D$ and $B$
meson decays at RHIC energy. Displayed are the spectra without
(''initial'') and with energy loss according to model II in 
\protect\cite{PRC98}.
}
\end{minipage}
\end{figure}

\section {Summary} 

In summary we analyze the beam energy dependence of various expected
sources of dileptons in ultra-relativistic heavy-ion collisions. Already
at $\sqrt{s} >$ 20 GeV a copious production of charm gives rise to
a dominant contribution to the dilepton spectrum at intermediate
invariant masses. Taking into account the ALICE detector acceptance we
study systematically the effect of single-electron transverse
momentum cuts. We find a threshold like behavior of the invariant mass
spectra of dileptons from primary correlated charm and bottom decays and
Drell-Yan yield as well: these sources are suppressed
at $M < M^{\rm min} \approx 2 p_\perp^{\rm min}$ for
$p_\perp^{\rm min} >$ 3 GeV. In contrast to this, the thermal dilepton signal
exhibits a plateau in this region which offers the opportunity to
identify them and to gain information on the initial stages of deconfined
matter at LHC energies.
The same mechanism also works at RHIC energies, however the expected
count rates are too small to make such a strategy feasible.
The complex decay chains of heavy mesons,
in particular open bottom, and the resulting
combinatorial background make an explicit identification of the
''hadronic cocktail'' very desirable to allow a safe
identification of the thermal signal
from deconfined matter. 

The selection of decay electrons from open charm and bottom can signalize
the presence of deconfined matter via a modified single-electron
$p_\perp$ spectrum in comparison with pp collisions.\\[9mm]
Stimulating discussions with 
P. Braun-Munzinger, Z. Lin, R. Vogt, G. Zinovjev are gratefully
acknowledged.
O.P.P. thanks for the warm hospitality of the nuclear theory group
in the Research Center Rossendorf.
The work is supported by BMBF grant 06DR829/1 and WTZ UKR-008-98.\\

\end{document}